\documentclass{aa}
\usepackage{txfonts}
\usepackage{graphicx}
\usepackage{t1enc}
\usepackage{epsfig}

\newcommand{\Td}    {T_\mathrm{d}}
\newcommand{\Tex}   {T_\mathrm{ex}}

\newcommand{\kms}   {km~s$^{-1}$}
\newcommand{\cmt}   {cm$^{-3}$}
\newcommand{\jpb}   {$\rm Jy~beam^{-1}$}	
\newcommand{\lo}    {$L_{\sun}$}
\newcommand{\mo}    {$M_{\sun}$}
\newcommand{\nh}    {NH$_3$}

\newcommand{\et}    {et al.}
\newcommand{\eg}    {e.\,g.,}
\newcommand{\uchii} {UC\ion{H}{ii}}

\begin{document}

\title{Unveiling the nature and interaction of the \\
intermediate/high-mass YSOs in IRAS~20343+4129\thanks{The
fits files for Figs.~\ref{fcont} and \ref{fchmaps} are also available in
electronic format at the CDS via anonymous ftp to cdsarc.u-strasbg.fr
(130.79.128.5) or via http://cdsweb.u-strasbg.fr/cgi-bin/qcat?J/A+A/.}}

\author{
Aina Palau$^{1,2}$, Robert Estalella$^2$, Paul T. P. Ho$^{3,4}$, Henrik
Beuther$^5$, \& Maria T. Beltr\'an$^2$}

\offprints{Aina Palau,\\ \email{apalau@laeff.inta.es}}

\institute{
Laboratorio de Astrof\'{\i}sica Espacial y F\'{\i}sica Fundamental, INTA, 
P.O. Box 78, E-28691 Villanueva de la Ca\~nada, Madrid, Spain 
\and
Departament d'Astronomia i Meteorologia, Universitat de Barcelona,
Av.\ Diagonal 647, E-08028 Barcelona, Catalunya, Spain
\and
Harvard-Smithsonian Center for Astrophysics,
60 Garden Street, Cambridge, MA 02138, USA
\and
Academia Sinica, Institute of Astronomy and Astrophysics, P.O. Box 23-141,
Taipei, 106, Taiwan
\and
Max-Planck-Institut for Astronomy, Koenigstuhl 17, 69117 Heidelberg, Germany
}

\date{Received / Accepted}

\authorrunning{Palau \et}
\titlerunning{
Star formation in IRAS~20343+4129
}

\abstract{IRAS~20343+4129 was suggested to harbor one of the most massive and
embedded stars in the Cygnus OB2 association, IRS~1, which seemed to be
associated with a north-south molecular outflow. However, the dust emission
peaks do not coincide with the position of IRS~1, but lie on either side of
another massive Young Stellar Object (YSO), IRS~3, which is associated with
centimeter emission.}{The goal of this work is to elucidate the nature of IRS~1
and IRS~3, and study their interactions with the surrounding medium.}{The
Submillimeter Array (SMA) was used to observe with high angular resolution the
1.3~mm continuum and CO\,(2--1) emission of the region, and we compared this
millimeter emission with the infrared emission from 2MASS.}{Faint millimeter
dust continuum emission was detected toward IRS~1, and we derived an associated
gas mass of $\sim0.8$~\mo. The IRS~1 Spectral Energy Distribution (SED) agrees
with IRS~1 being an intermediate-mass Class I source of about 1000~\lo, whose
circumstellar material is producing the observed large infrared excess. We have
discovered a high-velocity CO\,(2--1) bipolar outflow in the east-west
direction, which is  clearly associated with IRS~1.  Its outflow parameters are
similar to those of intermediate-mass YSOs. Associated with the blue large-scale
CO\,(2--1) outflow lobe, detected with single-dish observations, we only found
two elongated low-velocity structures on either side of IRS~3. The large-scale
outflow lobe is almost completely resolved out by the SMA. Our detected
low-velocity CO structures are coincident with elongated H$_2$ emission
features. The strongest millimeter continuum condensations in the region are
found on either side of IRS~3, where the infrared emission is extremely weak.
The CO and H$_2$ elongated structures follow the border of the millimeter
continuum emission that is facing IRS~3. All these results suggest that the dust
is associated with the walls of an expanding cavity driven by IRS~3, estimated
to be a B2 star from both the centimeter and the infrared continuum
emission.}{IRS~1 seems to be an intermediate-mass Class I YSO driving a
molecular outflow in the east-west direction, while IRS~3 is most likely a more
evolved intermediate/high-mass star that is driving a cavity and accumulating
dust in its walls. Within and beyond the expanding cavity, the millimeter
continuum sources can be sites of future low-mass star formation.}




\maketitle

\section{Introduction}

On the northeastern side of the Cygnus OB2 association, and at 1.4~kpc of
distance from the Sun (Le Duigou \& Kn\"odlseder 2002; Sridharan \et\ 2002)
there is a rimmed feature bright at centimeter wavelengths (Carral \et \ 1999)
and in the H$_2$ emission line at 2.12 $\mu$m (Kumar \et\ 2002), which  harbors at
its center the source IRAS~20343+4129. This IRAS source is a high-mass protostar
candidate of 3200~\lo\ (Sridharan \et\ 2002) embedded in dense gas (Richards \et
1987; Miralles \et\ 1994; Fuller \et\ 2005; Fontani \et\ 2006). When observed
with high angular resolution, two bright nebulous stars, IRS~1 (north) and IRS~3
(south),  are found inside the IRAS error ellipse (Kumar \et \ 2002). Either or 
both sources might account for the bulk of the total
luminosity of the IRAS source. Comer\'on \et\ (2002) carried out a study of the
red and massive objects in the entire Cygnus~OB2 association and conclude that
IRS~1 may be one of the most luminous and deeply embedded members of the OB
association, based on 2MASS $JHK$ photometry and spectroscopy in the range
1.5--2.4~$\mu$m. Extending southward from IRS~1 and surrounding IRS~3, there is
H$_2$ line emission in a fan-shaped structure which corresponds very well with a
blueshifted CO\,(2--1) lobe detected in single-dish observations by Beuther \et\ (2002b). The
CO structure found by these authors is bipolar, and the redshifted lobe is
centered on IRS~1, suggestive of a north-south molecular outflow. Kumar \et\
(2002) detect extended H$_2$ emission in the east-west direction toward IRS~1,
and attribute this emission to arise in a circumstellar disk, perpendicular to
the north-south outflow. Although all these observations seem to indicate that
IRS~1 is a high-mass YSO, no significant amount of ionized gas is found
associated with IRS~1 (Carral \et\ 1999). The only compact centimeter continuum source in
the region is associated with IRS~3, which is interpreted as an Ultra Compact
\ion{H}{ii} (\uchii) region ionized by a B2 star (Miralles \et\ 1994; Carral
\et\ 1999). Single-dish observations at 1.3~mm reveal two millimeter continuum peaks lying
on either side of IRS~3 (Beuther \et \ 2002a; Williams \et \ 2004). Thus there
is no clear evidence on whether IRS~1 and/or IRS~3 is the infrared source
producing most of the IRAS luminosity, and their relation with the
dust condensations and the outflow emission remains unclear.



In this paper we present SMA observations of the continuum emission at 1.3~mm and the CO\,(2--1)
emission toward IRAS~20343+4129. Our data provide an angular resolution sixteen times better in
area than that of the single-dish observations, allowing us to gain insight into the nature of
each source in the region.

\section{Observations}

The SMA\footnote{The Submillimeter Array is a joint project between the
Smithsonian Astrophysical Observatory and the Academia Sinica Institute of
Astronomy and Astrophysics, and is funded by the Smithsonian Institution and the
Academia Sinica.} (Ho \et \ 2004) was used to observe the 1.3~mm continuum
emission and the CO\,(2--1) (230.53796~GHz) emission toward IRAS~20343+4129. The
observations were carried out on 2003 August 3, with 6 antennas in the array. We
note that this was one of the first days for the SMA to work with 6 antennas.
The phase center was  $\alpha(\mathrm J2000)=20^{\mathrm h}36^{\mathrm
m}07\fs3$, $\delta(\mathrm J2000)=+41\degr39\arcmin57\farcs20$, and the
projected baselines ranged from 13.1 to 119.8~m. The pads of the antennas were
1, 4, 5, 8, 11, and 16, which correspond to a hybrid between the
compact and extended configurations.  System temperatures were around 200~K.  The full
bandwidth for each sideband at that time was 0.984~GHz, and each sideband was
divided into three blocks with four basebands in each block. The correlator was set
to the standard mode, which provided a spectral resolution of 0.8125~MHz (or
1.06 \kms\ per channel) across the full bandwidth of 1~GHz. The FWHM of the
primary beam at 230 GHz was $\sim\!45''$.



The raw visibility data were flagged and calibrated with the
MIR-IDL\footnote{The MIR cookbook by C. Qi can be found at
http://cfa-www.harvard.edu/~cqi/mircook.html.} package. The passband response was
obtained from observations of Uranus, which provided flat baselines when applied
to Neptune. The baseline-based calibration of the amplitudes and phases was
performed by using the source 2015+371. Typical rms of the phases was
$\sim\!60\degr$, yielding a positional uncertainty of $\sim\!0\farcs5$. Flux
calibration was set by using Uranus,  and the uncertainty in the absolute flux
density scale was $\sim\!20$\%.  





Imaging was conducted using the standard procedures in AIPS (for the line), and
MIRIAD (for the continuum; Sault \et\ 1995). The channel maps were cleaned
adding a value for the zerospacing parameter of 100~Jy (estimated from the
single-dish observations of Beuther \et\ 2002b) in the IMAGR task of AIPS, and
using different clean boxes. The continuum map was made using only the lower sideband
and excluding the CO\,(2--1) line (the upper sideband was noisier and combining
both sidebands did not result in a higher S/N).  We used three clean boxes to clean
the continuum, one box for each condensation detected with single-dish, and the
other box toward IRS 1. Cleaning with these boxes minimized the 
negative sidelobes in the final cleaned map. The final synthesized beam is
$3\farcs47 \times 2\farcs64$, with P.A.$=-38\fdg0$, the rms noise of the continuum
map is 2~m\jpb, and the rms noise of the 2.11~\kms\ wide channel maps is
0.4~\jpb.


\begin{table}
\caption{Parameters of the sources detected above 5$\sigma$ at 1.3~mm in the
IRAS~20343+4129 region}
\begin{center}
{\small
\begin{tabular}{lccccc}
\noalign{\smallskip}
\hline\noalign{\smallskip}
&\multicolumn{2}{c}{Position$^\mathrm{a}$}
&$I_\mathrm{\nu}^\mathrm{peak}$~$^\mathrm{b}$
&$S_\mathrm{\nu}^\mathrm{b}$
&Mass$^\mathrm{c}$
\\
\cline{2-3}
Source
&$\alpha (\rm J2000)$
&$\delta (\rm J2000)$
&(m\jpb)
&(mJy)
&(\mo)\\
\noalign{\smallskip}
\hline\noalign{\smallskip}
MM1   		&20:36:05.56 	&+41:40:00.2	&26.8	&44.8	&1.2\\
MM2   		&20:36:06.30 	&+41:40:00.7	&40.5	&40.5	&1.0\\
MM3       	&20:36:06.31	&+41:39:56.4	&29.9  	&34.0   &0.9\\
MM4		&20:36:06.62 	&+41:40:00.6	&39.7  	&45.3   &1.2\\
MM5     	&20:36:07.49 	&+41:40:12.8	&25.8  	&34.8   &0.9\\
MM6$^\mathrm{d}$&20:36:07.56 	&+41:40:08.0	&16.9	&31.5   &0.8\\
MM7		&20:36:08.19 	&+41:39:54.8	&16.7	&28.0   &0.7\\
\hline
\end{tabular}
\begin{list}{}{}
\item[$^\mathrm{a}$] Positions corresponding to the intensity peak.
\item[$^\mathrm{b}$] Corrected for the primary beam response. 
\item[$^\mathrm{c}$] Masses derived assuming a  dust temperature of 30~K, and a
dust mass opacity coefficient from Ossenkopf \& Henning (1994, see main text). The
uncertainty in the masses due to the opacity law is estimated to be a factor of
4.
\item[$^\mathrm{d}$] Associated with IRS~1.
\end{list}
}
\end{center}
\label{tcont}
\end{table}

\section{Results}

\subsection{Continuum \label{srcont}}

\begin{figure*}
\begin{center}
\includegraphics[width=13.5cm]{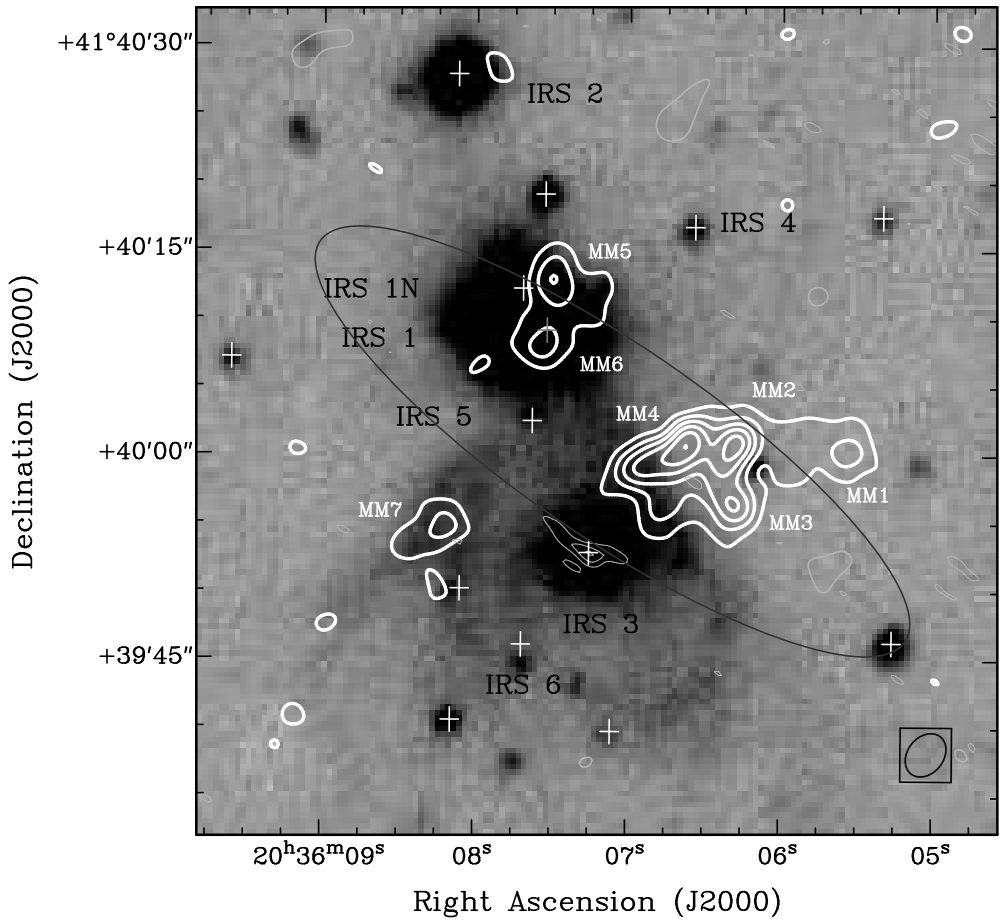}
\caption{
White (positive) and thick grey (negative) contours: SMA continuum emission at
1.3~mm towards the IRAS~20343+4129 region, obtained with natural weighting.
Contours are $-3$, 3, 6, 9, 12, 15, and 18 times the rms noise,
2~mJy~beam$^{-1}$. The synthesized beam, shown in the bottom right corner,  is  
$3\farcs5\times2\farcs6$, at $\mathrm{P.A.}=-38\fdg0$. Thin grey contours:
3.6~cm emission obtained with the VLA (Sridharan \et\ 2002). Contours are 3, 6,
and 9 times the rms noise, 0.2~mJy~beam$^{-1}$. Grey scale: H$_2$ emission
(continuum plus line) at 2.12 $\mu$m from Kumar \et \  (2002). The crosses
correspond to infrared sources from the 2MASS Point Source Catalog (PSC).
\label{fcont}
}
\end{center}
\end{figure*}


\begin{figure*}
\begin{center}
\includegraphics[width=14.5cm]{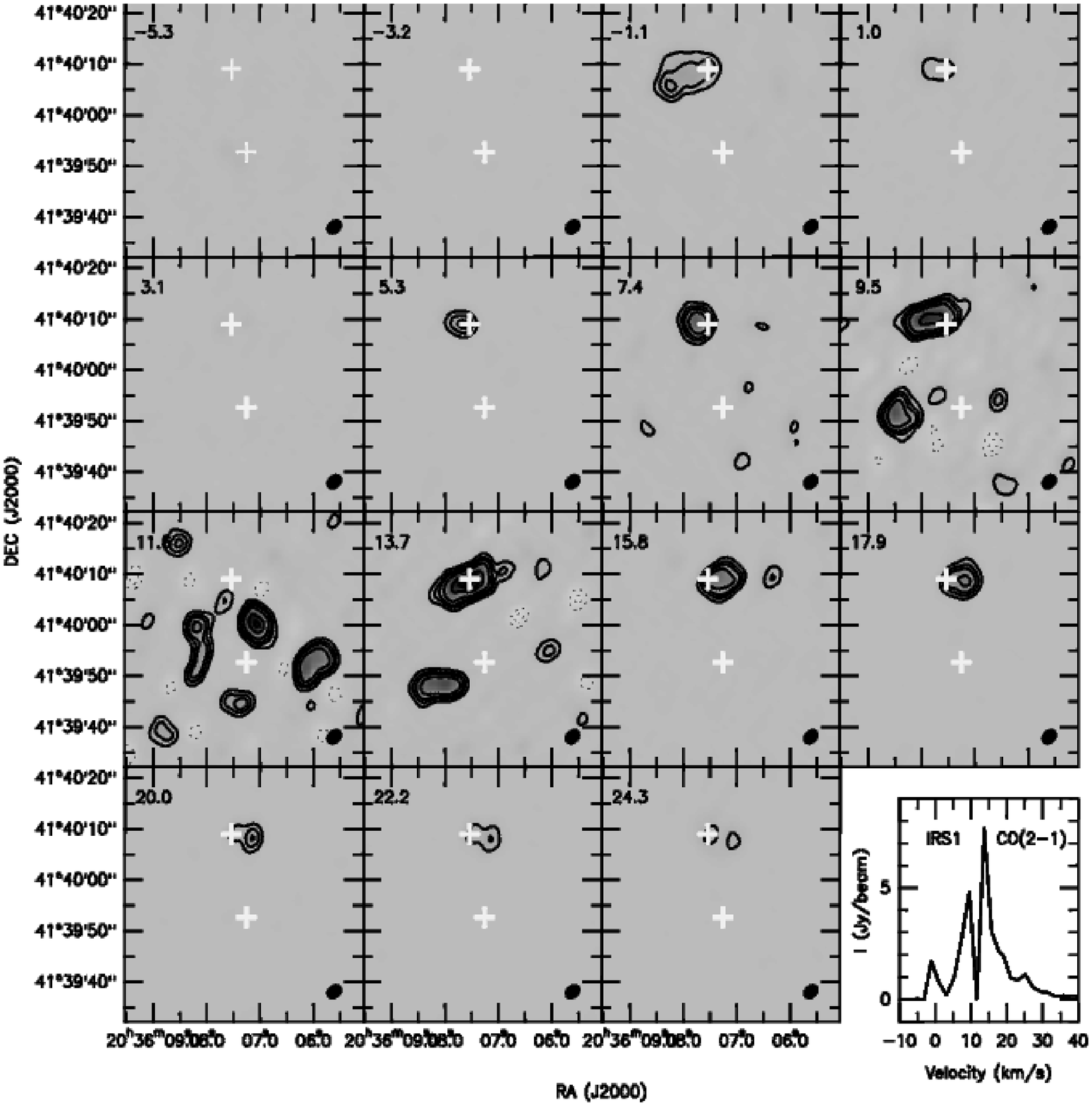}
\caption{
CO\,(2--1) channel maps of the IRAS~20343+4129 region, averaged over 2.11~\kms \
wide velocity intervals. The central velocity of each channel is indicated in
the upper left corner, and the systemic velocity is 11.5~\kms. The crosses
indicate the position of IRS~1 (north) and IRS~3 (south). The synthesized beam,
shown in the lower right corner, is $3\farcs5\times2\farcs7$, at
P.A.$=-37\fdg8$. Contours are $-6$, $-3$, 3, 6, 9, 15, 30, and 40 times the 
rms noise, 0.4~Jy~beam$^{-1}$. The lower right panel is the spectrum of the
CO\,(2--1) emission toward the position of IRS~1. The conversion factor is
2.46~K~(Jy/beam)$^{-1}$ (in the Rayleigh-Jeans assumption).
\label{fchmaps}
}
\end{center}
\end{figure*}

Figure~\ref{fcont} shows the continuum emission observed with the SMA at 1.3~mm.
The emission is found basically toward three positions in the field: to the west
of IRS~3 (where we found the strongest condensations, detected up to
18$\sigma$), to the east, and to the north of IRS~3 (where we found
condensations up to 9$\sigma$). The condensations to the west and to the east of
IRS~3 are coincident with the single-dish peaks of emission (Beuther \et\
2002a), and are not associated with any infrared source.  The emission to the
west of IRS~3 contains four condensations, MM1 to MM4, with a total flux density
of 230 mJy (corrected for the primary beam  response), while the emission to the
east has one condensation, MM7, of only 28~mJy.  This is different from the
single-dish measurements, for which the eastern condensation is stronger than
the western condensation  by almost a factor of 2. We estimated that the SMA has
picked up only 2\% of the flux density of the single-dish eastern
condensation (of $\sim\!30''$ in size), and 30\% of the flux density of the
single-dish western condensation (of $\sim\!20''$). Thus, the emission of the
eastern condensation is essentially extended, and has been filtered out by the
SMA, while the emission in the western condensation consists of different
compact millimeter sources.


In Table~\ref{tcont} we list the position, peak intensity, flux density and mass
for each millimeter continuum condensation detected above 5$\sigma$. In the
derivation of the masses, we assumed that all of the continuum emission is dust
emission which is optically thin, and adopted a gas-to-dust mass ratio of 100
and a dust mass opacity coefficient at 1.3~mm of 0.9~cm$^2$~g$^{-1}$
(agglomerated grains with thin ice mantles  in protostellar cores of densities
$\sim 10^6$~\cmt; Ossenkopf \& Henning 1994).  There is a factor of 4 in the
uncertainty of the masses due to uncertainties in the opacity law. As for the
dust temperature, we found two estimates in the literature. From \nh\
observations toward this region, Miralles \et\ (1994) derive a rotational
temperature of  $\sim\!20$ K, which can be considered a lower limit for the dust
temperature, as gas from protostellar envelopes is mainly heated by
collisions with warm dust grains (\eg\ Ceccarelli \et\ 1996). On the other
hand, by fitting two greybodies to the spectral energy distribution of
IRAS~20343+4129, Sridharan \et\ (2002) obtain a dust temperature $\Td$ for the
cold component of 44~K. However, in this last estimate of $\Td$, the flux
densities were measured with single-dish telescopes with angular resolutions
between 10$''$ and 100$''$, including the contribution from IRS~1, IRS~3 and
other sources in the region. Thus, $\Td=44$~K can be considered an upper limit.
We adopt the intermediate value of $\Td=30$~K.  

The mass of MM7 is $\sim\!0.7$~\mo, and the total mass of MM1 to MM4 is
$\sim\!6$~\mo. As a reference, the masses derived for the eastern and western
single-dish condensations are 44 and 23~\mo, respectively (Beuther \et\ 2002a,
2005; the authors adopt a dust temperature and an opacity law that yield masses
very similar to those obtained using our assumptions). Thus, the millimeter
compact sources detected with the SMA are embedded in a more massive gas halo.
For MM6, the dust condensation associated with IRS~1, we obtained a mass of
$\sim 0.8$~\mo. There is one condensation, MM5, $5''$ to the north of IRS~1 that
is slightly offset ($\sim\!2''$) to the west of the infrared source IRS~1N
(Fig.~\ref{fcont}). Since the SMA and 2MASS positional uncertainties are
$0\farcs5$ and $0\farcs6$ (Skrutskie \et\ 2006), respectively,  it is not clear
from our data whether MM5 is associated with IRS~1N or is tracing a different
source. Finally, we did not detect IRS~3 at 1.3~mm, setting an upper limit for
its cirmcumstellar mass of $\sim\!0.2$~\mo. It is worth noting that IRS~3 is the
only source associated with centimeter continnum emission in the field (see
Fig.~\ref{fcont}, and \S~\ref{sdirs3}).





\subsection{CO\,(2--1) \label{srco}}

Channel maps of the CO\,(2--1) emission are displayed in Fig.~\ref{fchmaps}.
CO\,(2--1) emission extends from $-2$ up to 33~\kms, with the systemic velocity being
11.5~\kms. The strongest CO\,(2--1) feature is associated with IRS~1, and spans
several channels for blueshifted and redshifted velocities, as can be seen also
in the spectrum of the CO\,(2--1) emission toward IRS~1 (Fig.~\ref{fchmaps}).

\begin{figure}
\begin{center}
\includegraphics[width=9cm]{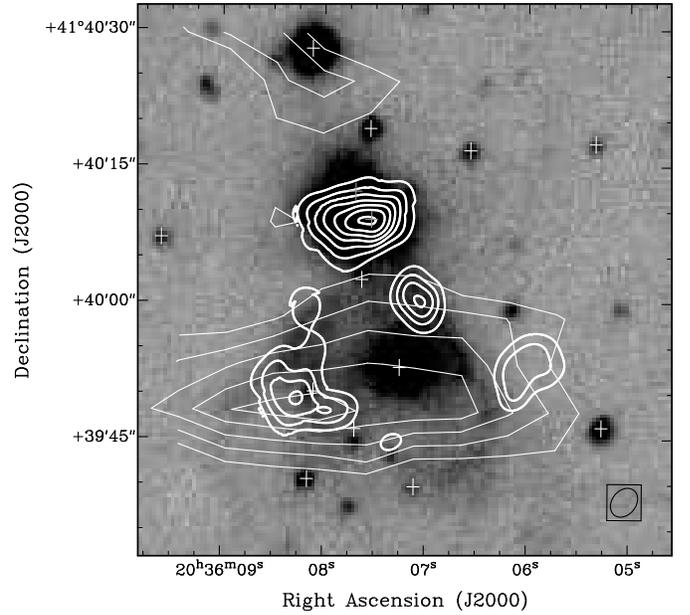}
\caption{
Thick contours: zero-order moment map for the low-velocity CO\,(2--1) emission.
Velocities have been integrated from 8.4 to 14.8~\kms. Contours range from 8 to
56~Jy~beam$^{-1}$~\kms, increasing in steps of 8~Jy~beam$^{-1}$~\kms. 
Thin contours: zero-order moment of the blueshifted CO\,(2--1) emission observed
in single-dish by Beuther \et\ (2002b). Contours are 5.2, 6.3, 7.4, 8.5, and
9.6~\jpb~\kms.
Grey scale: H$_2$ emission (continuum plus line) at 2.12 $\mu$m from Kumar \et
\  (2002). Note that the SMA low-velocity structures on either side of IRS~3 are
associated with H$_2$ extended emission.  The crosses correspond to infrared
sources from the 2MASS Point Source Catalog (PSC).
\label{fco21lv}
}
\end{center}
\end{figure}

The map of the low-velocity emission, integrated from 8.4 to 14.8~\kms, is
shown in Fig.~\ref{fco21lv}. From the figure, we can see clearly the association
of CO\,(2--1) with IRS~1, as well as the presence of low-velocity components
southwards of IRS~1 and surrounding IRS~3 that are associated with the
large-scale CO(2--1) blueshifted lobe from Beuther \et\ (2002b). It is worth
noting that the two elongated structures on either side of IRS~3 seem to be
associated with the fan-shaped structure found in  H$_2$ by Kumar \et \ (2002). 

\begin{figure*}
\begin{center}
\includegraphics[width=12cm]{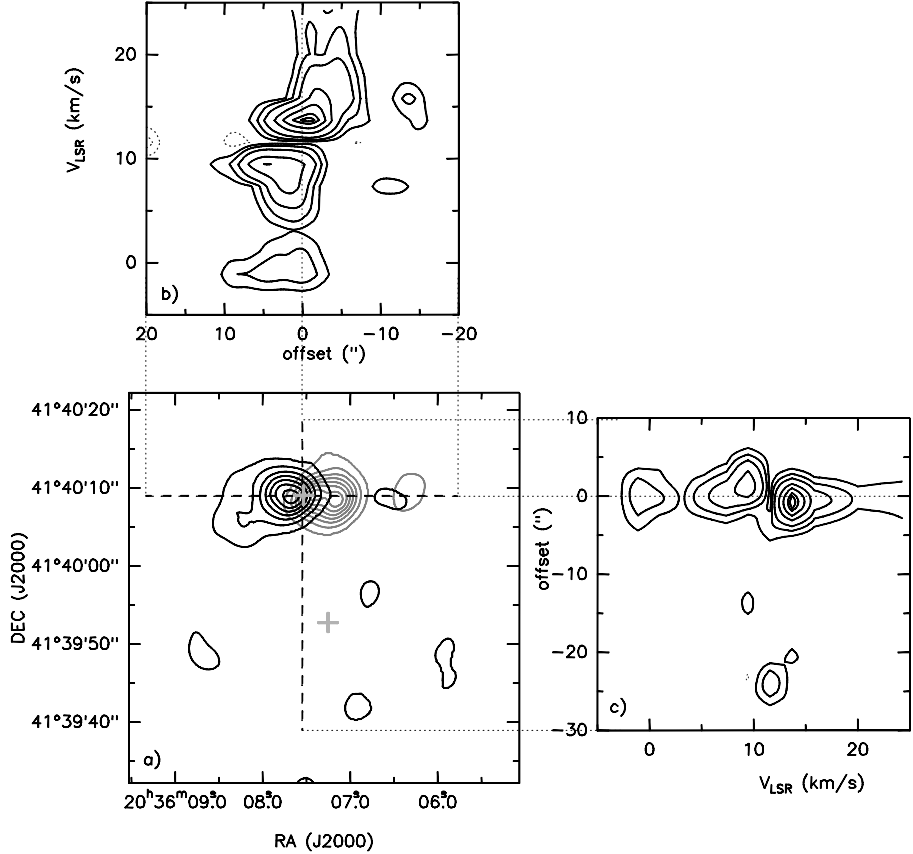}
\caption{
{\bf a)} CO\,(2--1) high-velocity emission toward IRAS~20343+4129. Light grey
contours, corresponding to the redshifted emission in a velocity range from 14.7
to 32.8~\kms, range from 2 to 38~Jy~beam$^{-1}$~\kms, increasing in steps of
6~Jy~beam$^{-1}$~\kms. Black contours, corresponding to the blueshifted emission
in a velocity range from $-6.4$ to 8.5~\kms, are the same as light grey
contours. Crosses mark the positions of IRS~1 (north) and IRS~3 (south).  
{\bf b)} Position-velocity (p-v) plot along the east-west direction centered on
IRS~1. Contours  are $-5$, $-2$, 2, 5, 10, 20, 30, 40, and 45 times
0.4~Jy~beam$^{-1}$.  
{\bf c)} The same as in b) along the north-south direction. 
\label{fco21hvpv}
} 
\end{center} 
\end{figure*}

Regarding the high velocities, these are only present in the immediate
surroundings of IRS~1. Figure~\ref{fco21hvpv}a plots the integrated
high-velocity emission toward IRS~1. Blue velocities have been integrated 
from $-6.4$ to 8.5~\kms, and red velocities from 14.7 to 32.8~\kms.
The high-velocity CO\,(2--1) emission has a bipolar structure, with the center
at the position of IRS~1, and is elongated in the east-west direction. Note that
the red lobe splits up into two subcomponents.

A position-velocity (p-v) plot obtained toward IRS~1 in the east-west direction
is shown in Fig.~\ref{fco21hvpv}b. Up to $\pm 8''$ from the zero offset
position, the emission shows a bipolar morphology, reaching high-velocities
that are blueshifted for positive offsets (to the east), and redshifted for
negative offsets (to the west). The distance from IRS~1 where we find high
velocities allows us to constrain whether such velocities are due to
gravitationally bound motions. We find weak high-velocity gas at
$\sim\!12$~\kms\ offset from the systemic velocity at position offsets up to $8''$, or
10000~AU. Such velocities at these distances would imply an extremely high
central mass of $\sim\!1000$~\mo\  for the motions to be gravitationally bound. Hence the
bipolar structure seen in CO\,(2--1) toward IRS 1 is most likely tracing an outflow motion.

Additionally, we computed the p-v plot toward IRS~1 in the north-south direction
(Fig.~\ref{fco21hvpv}c). The only clear CO feature along the direction of the
cut is the clump at the offset position zero, and does not seem related to any
other CO feature to the south of IRS~1, although one would expect a north-south
bipolar structure judging from the single-dish CO map from Beuther \et\ (2002b).
The CO emission at position zero arises from IRS~1, with the high-velocity
component coming from outflow motions (see above). Note that  observing outflow
emission in both the east-west (Fig.~\ref{fco21hvpv}b) and north-south
(Fig.~\ref{fco21hvpv}c) directions could be indicating that the IRS~1 outflow is
a wide-angle outflow (this could be also an effect of not resolving the base of
the outflow, but the high-velocity emission of Fig.~\ref{fco21hvpv}c, specially
in the blueshifted lobe, is partially resolved). Finally, the decrease in CO
emission at around the systemic velocity near the zero offset position, also
seen in the CO spectrum of Fig.~\ref{fchmaps}, may be produced by a combination
of the missing short spacings and opacity effects. Given that the brightness temperature at the line peak is
around 20~K, similar to the kinetic temperature, the line is optically thick at
systemic velocities. The CO emission at these velocities is probably
self-absorbed by foreground quiescent material of the cloud in which the YSOs
are embedded.  



We calculated the energetics of the outflow associated with IRS~1 for each lobe
separately, and listed the values in Table~\ref{toutpar}.  The expression used
for calculating the outflow CO column density $N(\mathrm{CO})$ from the
transition $J \rightarrow J-1$ (derived from Eq.~A1 of Scoville \et\ 1986) is:

{\footnotesize
\begin{displaymath}
\left[\frac{N(\mathrm{CO})}{\mathrm{cm^{-2}}}\right]=4.33 \times 10^{13}
\frac{\Tex}{J^2}\,
\mathrm{exp}\left(\frac{2.77J(J+1)}{\Tex}\right)
\frac{\tau_0}{1-e^{-\tau_0}}
\left[\frac{\int T_\mathrm{B}(v)dv}{\mathrm{K~km~s^{-1}}}\right],
\end{displaymath}
{\normalsize where $\Tex$ is the excitation temperature, $\tau_0$ is the optical
depth, and $T_\mathrm{B}(v)$ is the brightness temperature profile.}
}

For the mass derived from CO, we adopted a mean molecular weight per H$_2$
molecule of 2.8, and a CO abundance $X(\mathrm{CO})=10^{-4}$ (Scoville \et\
1986):

\begin{displaymath}
\left[\frac{M}{M_{\odot}}\right]=2.25 \times 10^{-16}
\left[\frac{A}{\mathrm{pc^2}}\right]
\left[\frac{N(\mathrm{CO})}{\mathrm{cm^{-2}}}\right],
\end{displaymath}
with $A$ being the area of the line emission.

We assumed optically thin emission in the line wing, and an excitation
temperature of $\sim\!25$~K, estimated from the spectrum in Fig.~\ref{fchmaps}.
Due to the lack of observations of other CO transitions, we could not make a
better estimate  of $\Tex$. However, this effect is small, as varying $\Tex$
between 15 and 30~K yields to a variation in the outflow parameters of only
$\sim\!7$\%. For the red lobe we integrated from 15 to 33~\kms, and for
the blue lobe from $-6$ to 8~\kms. The age or dynamical timescale
$t_\mathrm{dyn}$ was derived by dividing the size of each lobe (from the first
contour shown in Fig.~\ref{fco21hvpv}a) by the maximum velocity reached in the
outflow with respect to the systemic velocity (21.5~\kms\  for the red
lobe, and 17.5~\kms\  for the blue lobe).  We did not correct for the
inclination angle, since this parameter is not well known. To apply this
correction, the velocity must be divided by $\sin\ i$, and the linear size of
the lobes must be divided by $\cos\ i$, with $i$ being the inclination angle
with respect to the plane of the sky.


\begin{table*}[t]
\caption{Physical parameters$^\mathrm{a}$ of the outflow driven by IRS1}
\begin{center}
\begin{tabular}{lcccccccc}
\noalign{\smallskip}
\hline\noalign{\smallskip}
&Age
&$N_{12}$
&Mass
&$\dot{M}$
&$P$
&$\dot{P}$
&$E_{\mathrm{kin}}$
&$L_{\mathrm{mech}}$\\
Lobe
&(yr)
&(cm$^{-2}$)
&(\mo)
&(\mo~yr$^{-1}$)
&(\mo~\kms)
&(\mo~\kms~yr$^{-1}$)
&(erg)
&(\lo)\\
\noalign{\smallskip}
\hline\noalign{\smallskip}
Red
&3100 
&2.7$\times10^{16}$ 
&0.028
&9.0$\times10^{-6}$
&0.50
&1.6$\times10^{-4}$
&9.0$\times10^{43}$
&0.20\\
Blue
&3800 
&2.6$\times10^{16}$ 
&0.027
&7.2$\times10^{-6}$
&0.38
&1.0$\times10^{-4}$
&5.3$\times10^{43}$
&0.09\\
\hline
\end{tabular}
\begin{list}{}{}
\item[$^\mathrm{a}$] The parameters were obtained as follows.
Age:\,$t_\mathrm{dyn}$ (see main text); mass-loss rate:
$\dot{M}=M/t_\mathrm{dyn}$; momentum: $P=M V_\mathrm{range}$
($V_\mathrm{range}$ is the range for which we integrated the emission for each
lobe, see main text);  momentum rate (or mechanical force): 
$\dot{P}=P/t_\mathrm{dyn}$; energy of the outflow:
$E=1/2~M~V_\mathrm{range}^2$; mechanical luminosity:
$L_\mathrm{mech}=E/t_\mathrm{dyn}$. 
\end{list}
\end{center}
\label{toutpar}
\end{table*}

\subsection{Infrared emission from 2MASS \label{sr2mass}}

\begin{figure*}[t]
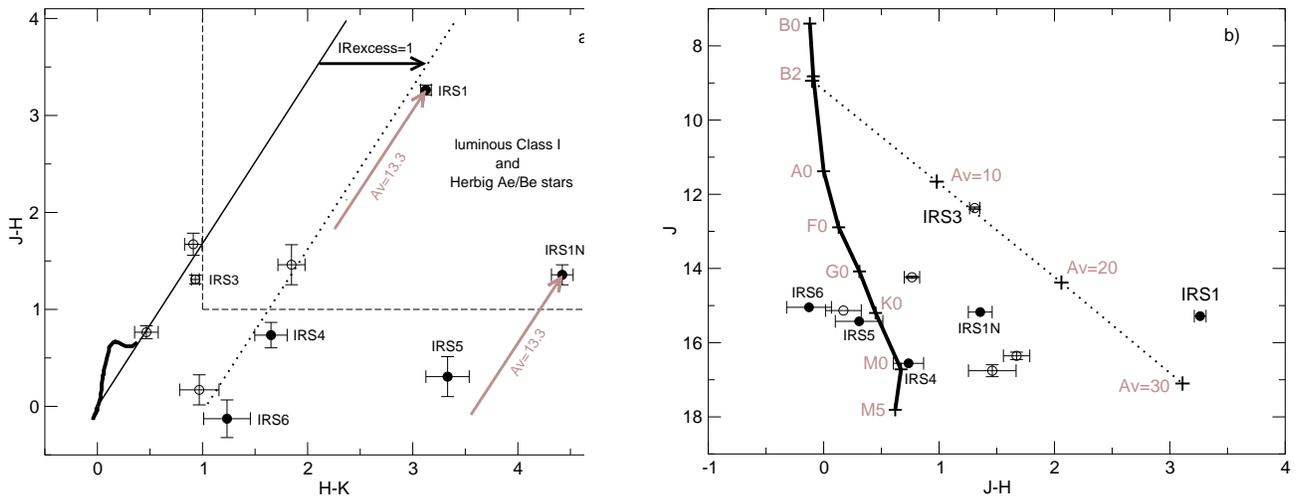

\begin{tabular}[b]{ll}
\epsfig{file=7692fi5a.eps, height=6.5cm} &
	\epsfig{file=7692fi5b.eps, height=6.5cm}\\
\end{tabular}
\caption{
{\bf a)} $(J-H)$, $(H-K)$ diagram of 2MASS sources within the SMA primary beam
toward the IRAS~20343+4129 region. The infrared excess is measured with respect
to the black thin solid line, which corresponds to the reddening line extending
from the loci of main-sequence stars, marked with a black thick solid line 
(Allen 1976, and extinction law of Rieke \& Lebofsky 1985). The dotted black
line indicates the locus of sources with infrared excess equal to one. Note that
IRS~3 has only a small infrared excess, while IRS~1 and IRS~1N are among the
sources with highest infrared excess. Thin long-dashed lines mark the region in
the diagram where most luminous Class I sources and Herbig Ae/Be stars are found
(Lada \& Adams 1992; Lee \et\ 2005).   
{\bf b)} $J$, $(J-H)$ diagram of the 2MASS sources in the IRAS~20343+4129
region. The solid black line indicates the locus of the main-sequence stars of
different spectral types (Allen 1976). Note that when dereddening IRS~3 along
the extinction vector (dotted line), it coincides with the position of B2
stars, for $A_\mathrm{V}=13.3$. In both panels, filled circles correspond to
the sources with infrared excess larger than 1.
}
\label{fcolors}
\end{figure*}

We extracted a sample of infrared sources within the SMA primary beam toward
IRAS~20343+4129 from the 2MASS Point Source Catalog (PSC, Skrutskie \et \ 2006),
with the aim of finding the possible infrared sources associated with the cloud
of gas and dust studied in this work, and plotted their $(J-H)$, $(H-K)$ diagram
(Fig.~\ref{fcolors}a).  In the diagram, there are three sources with low values
of the color indices, including IRS~3. These are unreddened (or only slightly
reddened) stars. We measured the infrared excess as the difference between the
$(H-K)$ color and the $(H-K)$ color correspoding to a reddened main-sequence
star. There is a group of five sources, listed in Table~\ref{t2mass}, for
which the infrared excess is larger than 1. We assume that such a large infrared
excess (typically, infrared excesses for Class II sources are smaller than 0.4;
Meyer \et\ 1997) is indicative of the YSOs being associated with the
IRAS~20343+4129 star-forming region.  Out of these five sources, we detected
dust continuum emission only toward IRS~1 and possibly IRS~1N.

\begin{table*}
\caption{2MASS sources within the SMA primary beam with infrared excess~$>1$}
\begin{center}
{\small
\begin{tabular}{lccccccc}
\noalign{\smallskip}
\hline\noalign{\smallskip}
&Identification
&\multicolumn{2}{c}{Position}
&
&
&
&infrared
\\
\cline{3-4}
Source
&2MASSJ+
&$\alpha (\rm J2000)$
&$\delta (\rm J2000)$
&$J$
&$H$
&$K$$^\mathrm{a}$
&excess$^\mathrm{b}$
\\
\noalign{\smallskip}
\hline\noalign{\smallskip}
IRS1   		&20360753+4140090  &20:36:07.53 &+41:40:09.1 &15.28 &12.02 &8.90  &1.17\\
IRS1N  		&20360769+4140121  &20:36:07.69 &+41:40:12.2 &15.18 &13.82 &9.40  &3.61\\
IRS3$^\mathrm{c}$&20360725+4139528 &20:36:07.25   &+41:39:52.8 &12.37 &11.06 &10.13 &0.15\\
IRS4		&20360656+4140167  &20:36:06.57 &+41:40:16.7 &16.56 &15.82 &14.17 &1.21\\
IRS5     	&20360762+4140024  &20:36:07.63 &+41:40:02.5 &15.43 &15.12 &11.79 &3.15\\
IRS6		&20360769+4139460  &20:36:07.69 &+41:39:46.1 &15.05 &15.17 &13.94 &1.31\\
\hline
\end{tabular}
\begin{list}{}{}
\item[$^\mathrm{a}$] The filter is $K_\mathrm{s}$, but we write $K$ for
simplicity.
\item[$^\mathrm{b}$] The infrared excess is measured as the difference between
the measured $(H-K)$ color and the $(H-K)$ color correspoding to a reddened
main-sequence star (Allen 1976, and extinction law of Rieke \& Lebofsky 1985). 
\item[$^\mathrm{c}$] Although IRS~3 does not have an infrared excess larger than 1, we
include this source in the table due to its relevance in the paper.
\end{list}
}
\end{center}
\label{t2mass}
\end{table*}

In order to estimate the spectral type of the infrared sources, we plotted them
in a $J$, $(J-H)$ diagram (Fig.~\ref{fcolors}b). In this diagram, if we
deredden IRS~3 along the extinction vector, it falls at the position of B2
stars,  consistent with the spectral type derived from  its centimeter continuum emission
(see \S~\ref{sdirs3} and Miralles \et\ 1994). Assuming that IRS~3 is a B2
main-sequence star, we derived the amount of visual extinction toward IRS~3,
$A_\mathrm{V}=13.3$~mag. Regarding IRS~4, IRS~5 and IRS~6, they all have 
spectral types around K0 or later.

In the $(J-H)$, $(H-K)$ diagram of Fig.~\ref{fcolors}a, when dereddening IRS~1
and IRS~1N by $A_\mathrm{V}=13.3$~mag (the visual extinction toward IRS~3), we
found that IRS~1 remains inside the loci of luminous Class I and Herbig Ae/Be
stars (Lada \& Adams 1992; Lee \et\ 2005), while IRS~1N, after dereddening, has
colors similar to YSOs of low luminosity.

\section{Discussion}

\subsection{The young high-velocity bipolar outflow toward IRS~1 \label{sdoutf}}

\paragraph{Small scale outflow found toward IRS 1:} 

The parameters of the IRS~1 outflow are similar to the mean values of low-mass
outflows (Wu \et\ 2004), and are about 2 orders of magnitude smaller than the
parameters derived from surveys of high-mass outflows (Beuther \et\ 2002b; Wu
\et\ 2005; Zhang \et\ 2005; Xu \et\ 2006). However, given that  the surveys of
high-mass molecular outflows are based on single-dish observations, picking up
large-scale structure and considering thus larger areas of outflow emission, we
compared the parameters of the IRS~1 outflow with other high-mass outflows
observed with interferometers, listed in Table~\ref{toutparcomp}. From the table
we found that the IRS~1 outflow is about 2--3 orders of magnitude less energetic
than the outflows of high-mass protostars observed with high angular resolution.
In some cases, the outflow parameters were corrected for opacity and
inclination, but these effects may contribute only about 1 order of magnitude.
From this comparison it seems that IRS~1 must be a low/intermediate-mass YSO. 


\begin{table*}[t]
\caption{Physical parameters of the IRS1 outflow compared with
other low, intermediate, and high-mass outflows observed with interferometers}
\begin{center}
{\small
\begin{tabular}{lccccccccc}
\noalign{\smallskip}
\hline\noalign{\smallskip}
&$L_\mathrm{bol}$
&Age
&Mass
&$\dot{M}$
&$P$
&$\dot{P}$
&$E_{\mathrm{kin}}$
&$L_{\mathrm{mech}}$\\
Region
&(\lo)
&(yr)
&(\mo)
&(\mo~yr$^{-1}$)
&(\mo~\kms)
&(\mo~\kms~yr$^{-1}$)
&(erg)
&(\lo)
&Ref.\\
\noalign{\smallskip}
\hline\noalign{\smallskip}
HH\,211
&3.6
&1400 
&0.0024
&1.7$\times10^{-6}$
&0.040
&2.8$\times10^{-5}$
&6.9$\times10^{42}$
&0.027
&1\\
I21391
&440
&$-$ 
&0.14
&$-$
&3.6
&1.4$\times10^{-3}$
&1.2$\times10^{45}$
&$-$
&2\\
\hline
I20343-IRS1
&3200
&3400 
&0.055
&1.6$\times10^{-5}$
&0.88
&2.6$\times10^{-4}$
&1.4$\times10^{44}$
&0.29
&3\\
\hline
I20293-A
&6300
&4300 
&2.0
&4.5$\times10^{-4}$
&90
&2.1$\times10^{-2}$
&4.1$\times10^{46}$
&79
&4\\
I18182
&20000
&$-$ 
&7.3
&$-$
&71
&$-$
&8$\times10^{45}$
&$-$
&5\\
ON2\,N
&$-$
&37000 
&58
&$-$
&1060
&2.8$\times10^{-3}$
&2.0$\times10^{47}$
&45
&6\\
\hline
\end{tabular}
\begin{list}{}{}
\item[REFERENCES:]  1: Palau \et\ (2006);  2: Beltr\'an \et\ (2002); 3: this
work; 4: Beuther \et\ (2004a); 5: Beuther \et\ (2006); 6: Shepherd \et\
(1997).
\end{list}
}
\end{center}
\label{toutparcomp}
\end{table*}

We additionally considered the centimeter continuum luminosity produced by shock
ionization that should be observed for the IRS~1 outflow. From the  correlation 
between the centimeter luminosity and the outflow momentum rate found by Anglada
(1995) for a sample of low/intermediate-mass YSOs, the IRS~1 outflow can account
for a centimeter luminosity of $\sim0.6$~mJy~kpc$^2$, assuming that all of the
stellar wind is shocked. This centimeter luminositiy is undetectable with the
sensitivity of the observations of Sridharan \et\ (2002, shown in
Fig.~\ref{fcont}), which is on the order of $0.4$~mJy~kpc$^2$. Then, the momentum rate
derived for the outflow of IRS~1 is not able to produce a detectable amount of
centimeter continuum emission, which is consistent with our observations.

Finally, the image of the H$_2$ line at 2.12~$\mu$m (1--0 S(1)) reveals strong 
emission very close to IRS~1 (Kumar \et\ 2002), being elongated in the east-west
direction, and thus coincident with the direction of the outflow of IRS~1. This
would suggest that the H$_2$ emission at 2.12~$\mu$m close to IRS~1 arises from
shocks in the outflow. However, Comer\'on \et\ (2002) detected line emission at
2.225~$\mu$m, which could be due to the 1--0~S(0) line of H$_2$. This H$_2$ line
at 2.225~$\mu$m has been found toward Class I and flat-spectrum sources (\eg\
Doppmann \et\ 2005), associated with outflows (Everett \et\ 1995; Davis \& Smith
1999; Caratti o Garatti \et\ 2006), and with photon-dissociated regions (Ramsay
\et\ 1993; Luhman \et\ 1998), but usually the H$_2$ line at 2.225~$\mu$m is much
fainter than the H$_2$ line at 2.12~$\mu$m, while this is clearly not the case
for IRS~1 (see Fig.~7 of Comer\'on \et\ 2002). Ratios of different intensities
of H$_2$ lines are used to discriminate between  shock-excited H$_2$ emission
and excitation by fluorescence. A detailed analysis of the ratios of high spectral
resolution observations of different H$_2$ lines would allow us to study the
mechanism of the H$_2$ excitation, but this is out of the scope of this paper.

\paragraph{Large-scale CO emission:}   Single-dish observations of CO\,(2--1)
toward IRAS 20343+4129 show a blue lobe centered around IRS~3, and a red
single-peaked lobe around IRS~1 (Beuther \et\ 2002b). H$_2$ emission in a
fan-shaped structure is found to be associated with the blue CO large-scale lobe
(see Fig.~\ref{fco21lv}).  This seems to suggest the existence of an outflow in
the north-south direction. However, from the SMA data we found no evidence of
such a north-south outflow (see Fig.~\ref{fco21hvpv}a, c). 

We compared the SMA CO\,(2--1) channel maps with the single-dish CO\,(2--1)
channel maps of H. Beuther. The blue lobe observed with the single-dish data
results from integrating only from 8 to 9~\kms, an interval very close to the
systemic velocity. In the SMA channel maps from Fig.~\ref{fchmaps}, we only
detected emission at the position of the blue single-dish lobe for velocities
between 8.4 and 14.8~\kms. Considering that the large-scale blue lobe arises
from extended emission of $\sim\!25''$ in size, we found that this emission
could not have been detected by the SMA. Given the shortest baseline of an
interferometer, and the size of the observed emission, one can estimate the
fraction of correlated flux detected by the interferometer. For a source of
$25''$, this corresponds to a 0.2\% for our SMA configuration (with a shortest
baseline of 11~k$\lambda$). Thus, out of the total CO\,(2--1) flux observed in
single-dish for the blue lobe, 1310~Jy~\kms, the SMA could pick up only
2.6~Jy~\kms, which should be detected at a few times the rms of the SMA channel
maps, as is the case (see channel at 9.5~\kms\ of Fig.~\ref{fchmaps}). We
concluded that almost all of the large-scale blueshifted lobe seen in
single-dish has been resolved out by the SMA. Note that in the single-dish
images of Beuther \et\ (2002b) there is a blue contour at the position of IRS~1,
suggesting that if blueshifted emission had been integrated for velocities
$<8$~\kms, the blue lobe of the IRS~1 outflow detected with the SMA would appear
in the single-dish observations as well.



Regarding the single-dish red lobe, this was obtained integrating from 13 to
15~\kms\ (Beuther \et\ 2002b).  We estimated the contribution of the SMA
redshifted emission to the single-dish redshifted emission. The SMA integrated
intensity of the red lobe for the same range of velocities as Beuther \et\
(2002b) is 160~Jy~\kms, obtained from the spectrum toward IRS~1, and assuming a
size of the lobe of $\sim 10''$.  Beuther \et \ (2002b) obtain a flux density of
1300~Jy~\kms, implying that the SMA redshifted emission can account for about
12\% of the redshifted emission detected with the single-dish data, a contribution about
2 orders of magnitude higher than that for the blueshifted large-scale lobe.
This suggests that the blue and red lobes seen in the single-dish data, come from material
at different spatial scales. 



\subsection{Is IRS~3 driving a cavity around it? \label{sdirs3}}

As seen above, the CO\,(2--1) emission from the single-dish data (Beuther \et\ 2002b)
shows a slightly blueshifted large-scale lobe that corresponds well with the fan-shaped
structure seen in H$_2$ emission around IRS~3 (see Fig.~\ref{fco21lv}). In
addition, from the SMA data of this work, we found low-velocity CO elongated
structures on either side of IRS~3, as well as dust condensations also on either
side of IRS~3. All these observational features seem to suggest that IRS~3  is
interacting with the surrounding medium and  producing a shell of circumstellar
gas expanding away from IRS~3. We search for any kinematical evidence of such an
expanding cavity in the low-velocity CO emission, but the strong missing flux,
and the presence of at least one bipolar outflow and other YSOs surrounding
IRS~3, make this search difficult. However, a preliminary study of the \nh\
emission in the region revealed dense gas surrounding IRS~3 whose kinematics
were consistent with an expanding shell with IRS~3 in its center (Palau \et, in
prep.). In the following, we consider whether IRS~3 is able to drive such a
cavity.

Given that the centimeter continuum emission associated with IRS~3 (of 1.8~mJy, Sridharan
\et\ 2002) is consistent with an ionizing B2 star (Panagia 1973), and that the
luminosity-to-mass ratio of typical B2 stars is high enough to allow the star to
push away the surrounding material by radiation pressure (given typical dust
opacities  of molecular clouds, Calvet \et\ 1991; Anglada \et\ 1995), the first
interpretation to explore is that the cavity is driven by the radiation pressure from
IRS~3. This leads to a scenario in which IRS~1 is a low/intermediate-mass YSO
accounting for a small fraction of the total bolometric luminosity of IRAS,
while IRS~3 is a high-mass star accounting for most of the bolometric
luminosity. However, the cavity could also be driven by a stellar wind from
IRS~3. We studied the scenario of a wind-driven cavity by following the model of
Anglada \et\ (1995). In this model, the centimeter continuum emission is assumed to trace
an ionized stellar wind. We adopted a spectral index for the wind of 0.6, and
estimated a mass loss rate of the ionized material following Beltr\'an \et\
(2001). By noting from the observations the radius of the cavity ($\sim\!10''$),
the velocity of its walls ($\sim\!2$~\kms), and the external pressure of the
ambient cloud ($\sim\!3.3$~\kms, from the line width of the CO\,(2--1) line in
the walls of the cavity), we obtained a maximum radius, density, and dynamical
timescale of the cavity  of 12$''$, 2300~\cmt, and 11000~yr, respectively, which
are reasonable values for the region. This leads to a scenario in which IRS~3 is
not necessarily the most massive source of the region, and then IRS~1 could
account for most of the IRAS luminosity.

\begin{table}[t]
\caption{Summary of the centimeter observations carried out toward IRAS~20343+4129}
\begin{center}
\begin{tabular}{ccccl}
\noalign{\smallskip}
\hline\noalign{\smallskip}
$\lambda$
&$S_\mathrm{\nu}$
&Observing
&$\theta_\mathrm{FWHM}$
\\
(cm)
&(mJy)
&year
&($''$)
&Refs.
\\
\noalign{\smallskip}
\hline\noalign{\smallskip}
6
&$1.1\pm0.2$ 
&1989 
&$5''$
&Miralles \et\ (1994)
\\
3.6
&$1.3\pm0.3$ 
&1994 
&$8''$
&Carral \et\ (1999)
\\
&$1.8\pm0.1$
&1998 
&$1''$
&Sridharan \et\ (2002)
\\
2
&$<1.0$$^\mathrm{a}$ 
&1989 
&$5''$
&Miralles \et\ (1994)
\\
0.7
&$<9$$^\mathrm{a}$ 
&2003 
&$1\farcs5$
&Menten \et, in prep.
\\
\hline
\end{tabular}
\begin{list}{}{}
\item[$^\mathrm{a}$] Upper limits at the 3$\sigma$ level.
\end{list}
\end{center}
\label{tcm}
\end{table}

Thus, in the interpretation of the centimeter continuum emission as either an \uchii\ region, or an
ionized stellar wind, IRS~3 can create a cavity of swept up material around it.
In order to distinguish between these two scenarios, it would be very useful to consider the spectral energy distribution in the centimeter range for IRS~3. In
addition to the 3.6~cm measurement from Sridharan \et\ (2002), shown in this
work, there are other observations at 6, 3.6, and 2 cm, listed in
Table~\ref{tcm}. The only simultaneous measurements are those at 6 and 2~cm,
which result in a spectral index $\le -0.1$, consistent with an optically thin
\uchii\ region.  However, the other observations at 3.6~cm  are not consistent
with this flat spectral index. With the measurements at 3.6 and 6~cm, the
spectral index ranges from 0.3 to 0.9 (depending on the epoch of the
observations at  3.6~cm). This spectral index is consistent with emission from
an ionized wind (\eg\ Panagia \& Felli 1975). Thus, from the current data it is
not clear whether the source is variable with time, and what is the value of the
spectral index of the centimeter source associated with IRS~3. New observations
at 6, 3.6, 2, and 1.3~cm toward IRS~3 would give insight into which mechanism, a
stellar wind or radiation pressure, is driving the cavity.



\subsection{On the nature of IRS~1  \label{sdirs1}}

From an analysis of $JHK$ color-magnitude diagrams including the high-mass stars
of the Cygnus OB2 association, Comer\'on \et\ (2002) propose that IRS~1 may be
one of the most luminous and embedded objects in the entire association. In the
$K$, $(H-K)$ diagram, IRS~1 is the reddest object of the association, and
dereddening along the extinction vector yields a very bright $K$ magnitude
compared with the other massive stars. However, this plot sets an upper limit to
the intrinsic brightness because in the $K$ band there may be some contribution
from circumstellar material. For this reason, Comer\'on \et\ (2002) use the $H$,
$(J-H)$ diagram, which is not seriously affected by circumstellar emission, and
find again that IRS~1 is among the brightest. Since the color-magnitude diagrams
were made by using the Second Incremental Release of the 2MASS Point Source
Catalog (PSC), and this release is by now obsolete, we redid the diagrams with
magnitudes from the current release of 2MASS~PSC, and found  the same values
except for the $J$ magnitude, yielding $(J-H)=3.26$ instead of 4.23 used by
Comer\'on \et\ (2002).  Thus, dereddening along the extinction vector in the
$H$, $(J-H)$ diagram does not set IRS~1 among the brightest members of the
association, but yields that IRS~1 must be still intrinsically brighter than
stars with spectral type B0, for which typical luminosities are around
25000~\lo\ (Panagia 1973), at least one order of magnitude higher than the
bolometric luminosity of IRAS~20343+4129.

If IRS~1 were a high-mass ZAMS star, one would expect to detect centimeter continuum
emission from ionized gas, and to detect the star at optical wavelengths. However, 
no centimeter continuum emission was detected above the 3$\sigma$ level of 0.6~m\jpb, and
IRS~1 does not appear in the POSSII plates. If IRS~1 were a high-mass protostar,
deeply embedded and still accreting most of its mass, one would expect to
observe a massive envelope, with a mass of the order of the accreted mass,
surrounding the protostar.  But we did not detect a massive envelope toward
IRS~1 from the millimeter continuum emission (the circumstellar mass was
$\sim\!0.8$~\mo, \S~\ref{srcont}), and the outflow parameters were comparable to
those of low/intermediate-mass protostars (\S~\ref{sdoutf}). Thus, a high-mass
nature for IRS~1 is not consistent with our observations. All this suggests that
IRS~1 cannot be considered a reddened stellar photosphere but a YSO with a cold
envelope, and that the estimation of its brightness from the magnitude-color
diagrams gives only an upper limit. Note however that the $JHK$ magnitudes of
IRS~1 cannot be accounted for by a low-mass YSO at the distance of the region.




At this point we consider whether the continuum emission is really tracing all
the dust surrounding IRS~1. We considered four different possibilities. First,
the SMA could be filtering out large-scale emission; we considered this option
and we ruled it out because the largest angular scale (FWHM) to which the SMA is
sensitive is $\sim\!8''$, larger than $\sim\!4''$, the observed size of the
millimeter source associated with IRS~1 (MM6). In addition, the CO emission
observed with the SMA toward IRS~1 shows that the SMA is really sensitive to
scales larger than the size of the dust condensation associated with IRS~1, and
that the dust emission is significantly more compact than the CO emission.
Second, the dust could be optically thick; however, this possibility yields an unrealistically low value for the dust temperature, given the flux density at 1.3~mm
and the deconvolved size for MM6 of $3''$. Third, the dust could be sublimated;
since dust sublimates at $\sim\!1500$~K (Whitney \et\ 2004), a
considerable amount of gas at such a high temperature should be detected at
optical wavelengths. Fourth, CO emission has been detected with no continuum
emission for a few YSOs; however, these objects are evolved (Class II/III)
low-mass systems (\eg\ Andrews \& Williams 2005; Takeuchi \& Lin 2005). Thus,
none of these possibilities is convincing for the case of IRS~1, hinting that
the mass traced by the millimeter continuum emission is most likely all the
circumstellar mass associated with IRS~1. 



Therefore, the current data suggest that IRS~1 is a low/intermediate-mass YSO.
In order to further constrain the mass and the evolutionary stage of IRS~1, we
plotted the spectral energy distribution (SED) compiled from 2MASS (corrected
for interstellar extinction, see \S~\ref{sr2mass}), MSX, IRAS, observations in
the submillimeter range carried out with SCUBA on the JCMT (Williams \et\ 2004),
and the SMA (this work). As the IRAS flux densities may have contribution from
both IRS~1 and IRS~3, they are only upper limits. The fluxes measured by SCUBA
are also upper limits because the single dish is picking up large-scale
emission, partially arising from IRS~3 and  from the dust condensations on
either side of IRS~3. The angular resolution of the MSX images allowed us to estimate
the flux density from IRS~1 by integrating the mid-infrared emission in an
aperture of $\sim\,15''$ of diameter around IRS~1. 

\begin{figure}[ht]
\begin{center}
\includegraphics[width=8cm]{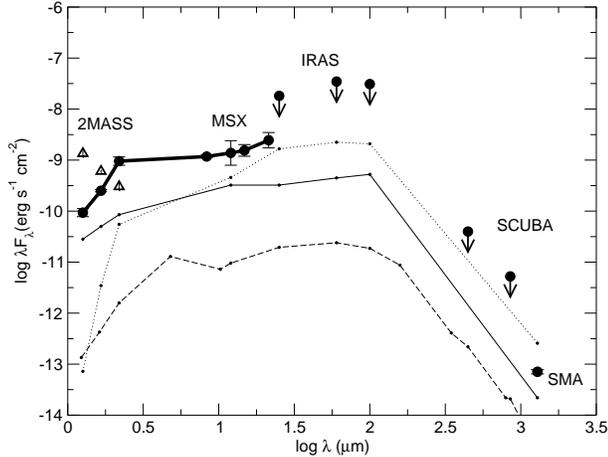}
\caption{\small
Spectral Energy Distribution (SED) for IRS~1. Black dots correspond to IRS~1, and thin curves
are SEDs from the literature for Class I sources of 960~\lo\ (dotted, VMR-D-IRS~13: Massi \et\
1999), 250~\lo\ (solid, VMR-D-IRS~14: Massi \et\ 1999), and 4~\lo\ (dashed, IRAS~04016+2610:
Eisner \et\ 2005), which have been scaled to the distance of IRAS~20343+4129. IRAS and SCUBA
fluxes are upper limits because it is not possible to disentangle the contribution of IRS~1
from the surrounding sources. For comparison, open triangles correspond to IRS~3 2MASS fluxes.
2MASS fluxes have been corrected for interstellar extinction (see
\S~\ref{sr2mass}).
\label{fsedirs1}
}
\end{center}
\end{figure}

The resulting SED (Fig.~\ref{fsedirs1}) shows a steep profile for the 2MASS
wavelengths, suggesting that any contribution from a hot photosphere is
negligible. Rather, the peak of the SED lying between 10 and 100~$\mu$m
indicates that most likely a cold envelope dominates the SED, with  d(log$\lambda
F_\lambda$)/d(log$\lambda$)$>0$ between 2 and 100 $\mu$m, consistent with the
classification of IRS~1 as a Class~I source (\eg\ Hartmann 1998). Note that if
IRS~1 were a Class 0 source, the SED should not show significant emission in the
near infrared (Andr\'e \et\ 1993; Lada 1999). We compared the IRS~1 SED with
SEDs of Class I sources of bolometric luminosities between 4 and 960~\lo\ from
the literature, scaled to the distance of IRAS~20343+4129. From this comparison,
we found that IRS~1 is likely not a low-mass source, but rather its SED
resembles that of intermediate-mass sources with bolometric luminosities around
1000~\lo\  (the exact value of the luminosity of IRS~1 cannot be determined
because the IRAS flux densities are only upper limits, and could be the main
contribution to the bolometric luminosity). From the evolutionary tracks of
Palla \& Stahler (1993) for intermediate-mass stars, and assuming that IRS~1 is
already in the birthline, one would expect a stellar mass for IRS~1 of around
5~\mo. This is a factor of 6 higher than the circumstellar mass derived from the
SMA data, $\sim\!0.8$~\mo\ (\S~\ref{srcont}). In the low-mass regime, Class 0
sources are expected to have circumstellar/envelope masses similar to the
stellar masses, and thus one could argue that $0.8$~\mo\ is too low for an
intermediate-mass YSO of 1000~\lo. However, one expects that in the Class I
stage, the ratio of the circumstellar-to-stellar mass progressively decreases.
Taking into account that intermediate-mass sources evolve faster to the
main-sequence than the low-mass sources, it may be reasonable to have a circumstellar mass lower than its
stellar mass. In addition, there are some cases in which the mass of the
dust/gas condensation associated with intermediate-mass YSOs is significantly
lower (at least by 1 order of magnitude) than the estimated stellar mass
(Beuther \et\ 2004b; Mart\'{\i}n-Pintado \et\ 2005; Zapata \et\ 2006). Thus, a
circumstellar mass of $\sim\!0.8$~\mo\ for IRS~1 seems to be compatible with its
classification as a Class I source of around 1000~\lo.

\subsection{Sources in different evolutionary stages: millimeter vs
near-infrared emission
\label{sdevolstage}}

In order to gain insight into the star formation process in the region, we
tentatively classified the different sources identified in this work in three
different evolutionary stages, depending on their millimeter and infrared
emission.

\paragraph{Sources at the end of the accretion phase:}    We classified in this
evolutionary stage IRS~3 to IRS~6 (Table~\ref{t2mass}), the sources detected only
in the infrared. From the centimeter and infrared emission, IRS~3 seems to be a
B2 star. As for IRS~4 to IRS~6, they all have spectral types around K0 or later
(as shown in Fig.~\ref{fcolors}b), and thus are low-mass YSOs. Since we did not
detect emission toward these sources at 1.3~mm, we estimated an upper limit for
their associated mass of $\sim\!0.2$~\mo\  (with the same assumptions as
\S~\ref{srcont}). These YSOs are possibly Class II/III sources. 

\paragraph{Sources in the main accretion phase:} We included in this group the
sources showing both infrared and millimeter emission, as is the case of IRS~1.
In \S~\ref{sr2mass} and \ref{sdirs1} we concluded that IRS~1 seems to be an
intermediate-mass Class I source.  Another source that could be in the main
accretion phase is IRS~1N, if we assume that it is associated with MM5 (see
\S~\ref{srcont}). For this source we estimated a spectral type A0 or later, from
the magnitude-color diagram of Fig.~\ref{fcolors}b, and an associated mass of
$\sim\!0.9$~\mo, from the SMA data. Thus, if IRS~1N is an embedded YSO, it is a
low-mass object. However, IRS~1N is the source with the highest $H-K$ color in
the region (see Fig.~\ref{fcolors}a), and a possibility for such a high $H-K$
color is that the $K$-band is contaminated by the H$_2$ line at 2.12~$\mu$m, as
the continuum-subtracted H$_2$ images from Kumar \et\ (2002) suggest.   Thus, it
remains unclear from the available data whether IRS~1N is a low-mass embedded
YSO, or an infrared source whose emission mainly arises in the interaction of an
outflow (either the IRS~1 outflow, or an outflow from IRS~1N/MM5 itself) with
the surrounding medium.


\paragraph{Starless core candidates:}  The sources MM1 to MM4 and MM7
(Table~\ref{tcont}) have been detected only in the millimeter, and lie in a
region completely dark in the near infrared (Fig.~\ref{fcont}). The masses of
these dust condensations are 0.7--1.2~\mo, and thus they are low-mass
condensations.  Sources bright only in the millimeter could also be tracing
Class 0 protostars being at the beginning/main accretion phase. However, given
that we did not find any sign of star formation such as outflow emission, 
we suggest that some, if not all, of these sources could be starless core
candidates.\\

Therefore, the IRAS~20343+4129 region harbors sources that seem to be in
different evolutionary stages, and with stellar masses ranging from $<0.2$ to
8--10~\mo.   The intermediate/high-mass sources of the region, IRS~1 and IRS~3,
are in different evolutionary stages. IRS~3 is visible at optical wavelengths,
does not show infrared excess,  and has no CO\,(2--1) nor dust emission
associated, suggesting that it has already finished the main accretion phase. On
the contrary, IRS~1 is not detected in the optical, has strong infrared excess
and dust emission associated, and is the driving source of a bipolar outflow,
which indicates that IRS~1 is still accreting a significant amount of mass, with
IRS~1 therefore younger than IRS~3. However,  even with different evolutionary
stages, IRS~1 and IRS~3 could have formed simultaneously, since the uncertainty
in their masses is important, and the contraction time to the main-sequence
(since the formation of a first hydrostatic core until the hydrogen burning
stage), which deacreases with the mass of the YSO, in the intermediate/high-mass
regime gets comparable or shorter than the free-fall timescale, and thus an
intermediate-mass Class I YSO may have formed simultaneously with a high-mass
object already in the main-sequence phase. For example, using the computations
of Bernasconi \& Maeder (1996), and assuming that IRS~3 is around 9~\mo\, and
IRS~1 is around ~5~\mo, IRS~3 has a contraction time of $\sim\!0.3$~Myr, and
IRS~1 of $\sim\!1$~Myr. This is different from the low-mass regime, where
YSOs evolve to the main-sequence in a timescale $>10$~Myr (\eg\ Hayashi 1961;
Iben 1965), which is always much larger than the free-fall timescale
($\sim0.5$~Myr). Thus, a low-mass YSO in the pre-main-sequence phase (Class
II/III) and a low-mass YSO in the main accretion phase (Class 0/I) cannot have
formed simultaneously. As seen in this section, we find low-mass YSOs in the
very first phases of formation, and others that seem to be Class II/III
candidates, and thus they cannot be coeval. Therefore, the star formation in
IRAS~20343+4129 seems to be a continuous process. It is worth noting that the
intermediate/high-mass star IRS~3 is in an evolutionary stage equal or more
evolved than the low-mass sources. This is in contrast to previous claims that
intermediate/high-mass stars in clusters appear less evolved than the low-mass
stars in the cluster (Massi \et\ 2000; Kumar \et\ 2006).

Finally, the spatial distribution of the sources found in IRAS~20343+4129 shows
that star formation in this region is localized, that is, to the north and south
of IRS~3 there are YSOs that are already bright in the infrared, while to the
east and to the west there are millimeter sources that seem to be starless
cores. Since IRS~3 and IRS~1 could be coeval, we suggest that these two sources
are reflecting the initial conditions of high density in the parental cloud. As
for the low-mass dust condensations on either side of IRS~3, we propose that
they may be compressed by the expanding cavity driven by IRS~3, and thus could
eventually form a new generation of stars.

\begin{figure}[ht]
\begin{center}
\includegraphics[width=9cm]{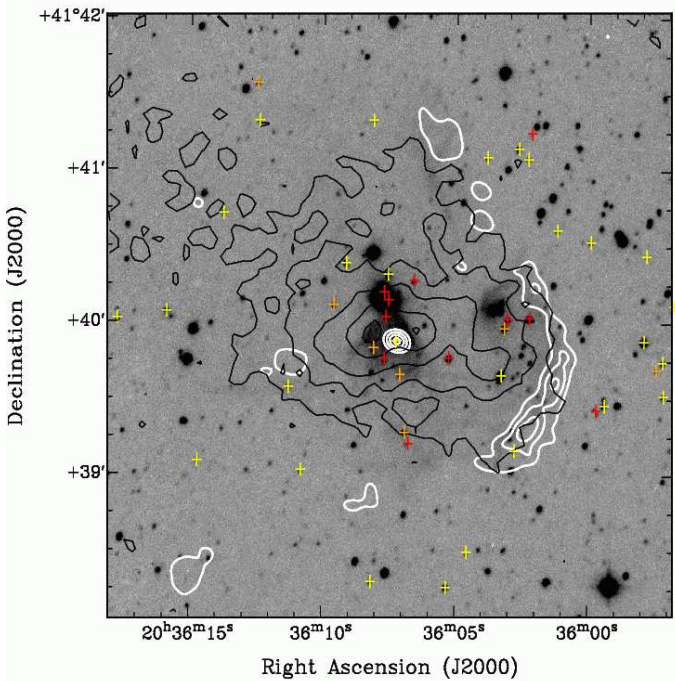}
\caption{Grey scale: H$_2$ emission (continuum plus line) at 2.12 $\mu$m in the
IRAS 20343+4129 region (Kumar \et\ 2002). White contours correspond to the
3.6~cm emission from Carral \et\ (1999), and are 0.66, 0.99, 1.14, 1.33, 1.52,
and 1.71~m\jpb.  Black contours are the 1.2~mm continuum emission observed in
single dish by Beuther \et\ (2002a), and trace the dust cloud where IRS~1 and
IRS~3 are forming. Contours are: 25, 50, 100, 200, and 300~m\jpb. Note that the
centimeter emission traces the ionized southwestern border of the cloud, as well
as IRS~3 in the center of the cloud. Crosses indicate the spatial distribution
of the 2MASS sources in the region with infrared excess larger than zero:
(0--0.4) for the yellow color (Class II sources); (0.4--1) for the orange color
(Class I), and $>1$ for the red color, following Matsuyanagi \et\ (2006). Note
that most of the 2MASS sources with infrared excess $>1$ lie in the center of
the cloud, near IRS~3 and IRS~1.
\label{fspatialdistr}
}
\end{center}
\end{figure}

\subsection{IRAS 20343+4129 within Cygnus OB2 \label{sdview}}

The H$_2$ emission at 2.12~$\mu$m shows a cometary arch about $1'$ to the
southwest of IRS~3, which is also detected in centimeter emission (see
Fig.~\ref{fspatialdistr}; Kumar \et\ 2002; Carral \et\ 1999). This cometary arch
follows the border of a dust cloud traced by 1.2~mm emission observed in
single-dish by Beuther \et\ (2002a), and could be produced by the ionization
front from a nearby OB star. In fact, the arch is facing the center of the
Cygnus OB2 association. This would be similar to the bright-rimmed clouds facing
HII regions, such as IC\,1396N (Sugitani \et\ 1991; Beltr\'an \et\ 2002), which
is ionized by an O6.5 star at 13~pc of distance (Schwartz \et\ 1991). We plotted
the OB stars of the association (Reed 2003\footnote{Catalog available at
http://othello.alma.edu/$\sim$reed/OBfiles.doc}), and found that there are five
O6--O9 stars about 25$'$ (9~pc at the adopted distance for IRAS~20343+4129) to
the south-west of the arch. Given the flux density of the centimeter continuum emission
tracing the cometary arch, one can estimate the required flux of ionizing
photons per unit area that must reach the cloud to produce the observed
centimeter emission (Lefloch \et\ 1997). In our case, the flux density of the
arch at 3.6~cm is around 5~mJy, and this requires a flux of ionizing photons per
area unit of $4\times10^9$~s$^{-1}$~cm$^{-2}$. The flux of ionizing photons for
an O6 star is tipically $1.20\times10^{49}$~s$^{-1}$ (Panagia 1973), and
assuming a distance to the arch of 9~pc, the flux of ionizing photons per area
unit reaching the cloud is $1.2\times10^9$~s$^{-1}$~cm$^{-2}$, close to the
value required to account for the flux of the centimeter emission in the arch.
Thus, the O6 star $25'$ to the southwest of the arch, BD+41\,3807, is probably
the star ionizing the border of the IRAS 20343+4129 cloud. Finally, we
calculated the infrared excess (as described in \S~\ref{sr2mass}) of the 2MASS
sources within a diameter of 4$'$ centered on IRS~3 (thus including a region
outside the rim), and plotted the sources with positive infrared excess in
Fig.~\ref{fspatialdistr}. In the figure, crosses with redder colors  correspond
to sources with larger infrared excess (and thus presumably younger). As seen
from the figure, most of the sources with strongest infrared excess within the
field are found inside the cloud, thus constituting a region of recent star
formation as compared to its surroundings. In addition, the millimeter sources
detected in this work could be sites of future star formation, and thus
IRAS~20343+4129 is a region actively forming stars within the Cygnus OB2
association.



\section{Conclusions}

We observed the dust continuum emission at 1.3~mm with the SMA as well as the
CO\,(2--1) emission toward the massive star-forming region IRAS~20343+4129, in
order to study the properties of the different protostars in the region and
their interactions with the surrounding medium.  Two bright infrared sources,
IRS~1 in the north and IRS~3 in the south, lie inside the SMA primary beam, and
IRS~3 is associated with centimeter continuum emission. Our main conclusions can be
summarized as follows:

\begin{enumerate}

\item The dust continuum emission reveals three main condensations, to the north
(associated with IRS~1), to the east and to the west of IRS~3, with the western
condensation being the brightest one and consisting of different 
subcondensations, of $\sim\!1$~\mo\ each. Toward the eastern condensation, of
$\sim\!0.7$~\mo, the SMA has filtered out most of the emission. The estimated
mass of the condensation associated with IRS~1 is $\sim\!0.8$~\mo.

\item We discovered a bipolar high-velocity CO outflow, elongated in the
east-west direction, and identified IRS~1 as its driving source. The millimeter
continuum and CO emissions indicate that IRS~1 is not a high-mass YSO, and the
SED agrees with IRS~1 being an intermediate-mass Class I source. 

\item The low-velocity CO emission shows, in addition to emission toward IRS~1,
two elongated structures on either side of IRS~3, coincident with 
extended H$_2$ emission at 2.12~$\mu$m. The emission from the blueshifted lobe
of the large-scale CO outflow seen in single-dish by Beuther \et\ (2002b) has
been filtered out by the SMA, and the lobe does not have compact emission. A
scenario in which the blue large-scale CO lobe traces a cavity blown up by
IRS~3, and where the dust condensations on either side of IRS~3 are the result
of the accumulation of mass in the walls of the expanding cavity, is consistent
with the observations.  In this scenario, the elongated low-velocity CO emission
and the H$_2$ extended emission trace the walls of the cavity. The expanding
cavity could be either driven by a stellar wind from IRS~3, or driven
by radiation if we assume that IRS~3 is a B2 star.

\item We found objects with very different properties and evolutionary stages
that have been born in the same parental cloud. In addition, these objects are
not randomly distributed in the cloud, but their distribution seems to be
determined, at least partially, by the accumulation of mass at the walls of an
expanding cavity driven by IRS~3.



\end{enumerate}

\begin{acknowledgements}

A.\ P. is deeply grateful to the SMA staff, Charlie Qi, Qizhou Zhang, and
\'Alvaro S\'anchez-Monge for assistance in the reduction process, as well as to
Stan Kurtz and M.\ S. Nanda Kumar for providing the centimeter and H$_2$ images.
Also thanks to Mayra Osorio and J. Miquel Girart for useful discussions,  and
Rosario L\'opez for technical support. A.\ P., R.\ E., and M.\ T.\ B. are
supported by a MEC grant AYA2005-08523 and FEDER funds. H.\ B. acknowledges
financial support by the Emmy-Noether-Program of the Dutsche
Forschungsgemeinschaft (DFG, grant BE2578). This publication makes use of data
products from the Two Micron All Sky Survey, which is a joint project of the
University of Massachusetts and the Infrared Processing and Analysis
Center/California Institute of Technology, funded by the National Aeronautics
and Space Administration (NASA) and the National Science Foundation. 


\end{acknowledgements}

{}


\begin{thebibliography}{}

\bibitem[]{}
Allen, C. W. 1976, \emph{Astrophysical Quantities}, 3rd edn., London:
Athlone Press

\bibitem[]{}
Andr\'e, P., Ward-Thompson, D., \& Barsony, M. 1993, ApJ, 406, 122

\bibitem[2005]{AW05}
Andrews, S.M., \& Williams, J.P. 2005, ApJ, 619, L175

\bibitem[]{}
Anglada, G. 1995, RMxAA, 1, 67

\bibitem[1995]{Anglada95}
Anglada, G., Estalella, R., Mauersberger, R., \et\ 1995, ApJ, 443, 682


\bibitem[]{}
Beltr\'an, M. T., Estalella, R., Anglada, G., Rodr\'{\i}guez, L. F., \&
Torrelles, J. M. 2001, AJ, 121, 1556


\bibitem[]{}
Beltr\'an, M. T., Girart, J. M., Estalella, R., Ho, P. T. P., \& Palau, A. 2002,
ApJ, 573, 246


\bibitem[]{}
Bernasconi, P. A., \& Maeder, A. 1996, A\&A, 307, 829

\bibitem[]{}
Beuther, H., Schilke, P., \& Gueth, F. 2004a, ApJ, 608, 330

\bibitem[2002a]{Beuther02a}
Beuther, H., Schilke, P., Menten, K.\ M., \et\ 2002a, ApJ, 566, 945 

\bibitem[]{}
Beuther, H., Schilke, P., Menten, K.\ M., \et\ 2005, ApJ, 633, 535 

\bibitem[2002b]{Beuther02b}
Beuther, H., Schilke, P., Sridharan, T. K., \et\ 2002b, A\&A, 383, 892  

\bibitem[]{} 
Beuther, H., Zhang, Q., Greenhill, L. J., \et\ 2004b, ApJ, 616, L31

\bibitem[]{}
Beuther, H., Zhang, Q., Sridharan, T. K., Lee, C.-F., \& Zapata, L. A. 2006,
A\&A, 454, 221

\bibitem[]{}
Bontemps, S., Andr\'e, P., Terebey, S., \& Cabrit, S. 1996, A\&A, 311, 858


\bibitem[]{}
Calvet, N., Pati\~no, A., Magris, G., \& D'Alessio, P. 1991, ApJ, 380, 617

\bibitem[]{}
Caratti o Garatti, A., Giannini, T., Nisini, B. \& Lorenzetti, D. 2006, A\&A,
449, 1077	

\bibitem[1999]{}
Carral, P., Kurtz, S., Rodr\'{\i}guez, L. F., \et\ 1999, RMxA\&A, 35, 97

\bibitem[1996]{}
Ceccarelli, C., Hollenbach, D. J., \& Tielens, A. G. G. 1996, ApJ, 471,
400


\bibitem[]{}
Comer\'on, F., Pasquali, A., Rodighiero, G., \et\  2002, A\&A, 389, 874

\bibitem[]{}
Davis, C. J., Smith, M. D., Eisl\"offel, J., \& Davies, J. K. 1999,
MNRAS, 308, 539



\bibitem[]{}
Doppmann, G. W., Greene, T. P., Covey, K. R., \& Lada, C. J. 2005, ApJ, 130,
1145

\bibitem[]{}
Eisner, J. A., Hillenbrand, L. A., \& Carpenter, J. M. 2005, ApJ, 635, 396

\bibitem[]{}
Everett, M. E., Depoy, D. L., Pogge, R. W. 1995, AJ, 110, 1295


\bibitem[]{}
Fontani, F., Caselli, P., Crapsi, A., \et\ 2006, A\&A, 460, 709

\bibitem[]{}
Fuller, G. A., Williams, S. J., \& Sridharan, T. K. 2005, A\&A, 442, 949


\bibitem[]{}
Hartmann, L. 1998, \emph{Accretion Processes in Star Formation}, eds. A.
King, D. Lin, S. Maran, J. Pringle, \& M. Ward, Cambridge: Cambridge University
Press, p. 8



\bibitem[]{}
Hayashi, C. 1961, PASJ, 13, 450

\bibitem[]{}
Ho, P.T.P., Moran, J.M., \& Lo, K.Y. 2004, ApJ, 616, L1

\bibitem[]{}
Iben, I., Jr. 1965, ApJ, 141, 993



\bibitem[]{}
Kumar, M. S. N., Bachiller, R., \& Davis, C.J. 2002, ApJ, 576, 313

\bibitem[]{}
Kumar, M. S. N., Keto, E., \& Clerkin, E. 2006, A\&A, 449, 1033


\bibitem[]{}
Lada, C. J. 1999, in \emph{The Origin of Stars and Planetary Systems}, eds. C.
J. Lada, and N. D. Kylafis, Kluwer Acad. Publ., p. 143

\bibitem[]{}
Lada, C. J., \& Adams, F. C. 1992, ApJ, 393, 278

\bibitem[]{}
Le Duigou, J.-M., \& Kn\"odlseder, J. 2002, A\&A, 392, 869

\bibitem[]{}
Lee, H.-T., Chen, W. P., Zhang, Z.-W., \& Hu, J.-Y. 2005, ApJ, 624, 808

\bibitem[]{}
Lefloch, B., Lazareff, B., \& Castets, A. 1997, A\&A, 324, 249

\bibitem[]{}
Luhman, K. L., Engelbracht, C. W., \& Luhman, M. L. 1998, ApJ, 499, 799

\bibitem[]{}
Mart\'{\i}n-Pintado, J., Jim\'enez-Serra, I., Rodr\'{\i}guez-Franco, A.,
Mart\'{\i}n, S., \& Thum, C. 2005, ApJ, 628, L61

\bibitem[]{}
Massi, F., Giannini, T., Lorenzetti, D., \et\ 1999, A\&AS, 136, 471

\bibitem[]{}
Massi, F., Lorenzetti, D., Giannini, T., \& Vitali, F. 2000, A\&A, 353, 598

\bibitem[]{}
Matsuyanagi, I., Itoh, Y., Sugitani, K., Oasa, Y., Mukai, T., \et\  2006, PASJ,
58, L29


\bibitem[]{}
Meyer, M. R., Calvet, N., \& Hillenbrand, L. A. 1997, AJ, 114, 288

\bibitem[]{}
Miralles, M. P., Rodr\'{\i}guez, L. F., \& Scalise, E. 1994, ApJSS, 92, 173 

\bibitem[]{} 	
Ossenkopf, V., \& Henning, Th. 1994, A\&A, 291, 943

\bibitem[]{} 	
Palau, A., Ho, P. T. P., Zhang, Q., \et\ 2006, ApJ, 636, L137

\bibitem[]{}
Palla, F., \& Stahler, S.W. 1993, ApJ, 418, 414


\bibitem[]{}
Panagia, N. 1973, AJ, 78, 929

\bibitem[]{}
Panagia, N., \& Felli, M. 1975, A\&A, 39, 1


\bibitem[]{}
Ramsay, S. K., Chrysostomou, A., Geballe, T. R., Brand, P. W. J. L., \& Mountain,
M. 1993, MNRAS, 263, 695

\bibitem[]{}
Reed, B. C. 2003, AJ, 125, 2531


\bibitem[]{} 
Richards, P. J., Little, L. T., Toriseva, M., \& Heaton, B.D. 1987, MNRAS, 228,
43 

\bibitem[]{}
Rieke, G. H., \& Lebofsky, M. J. 1985, ApJ, 288, 618


\bibitem[]{}
Sault, R. J., Teuben, P. J., \& Wright, M. C. H. 1995, in ASP Conf. Ser. 77,
Astronomical Data Analysis Software and Systems IV, ed. R. A. Shaw, H. E.
Payne, \& J. J. E. Hayes, San Francisco: ASP, p. 433


\bibitem[]{} 	 
Scoville, N. Z., Sargent, A. I., Sanders, D. B., \et\ 1986, ApJ, 303, 416

\bibitem[]{}
Shepherd, D. S., Churchwell, E., \& Wilner, D. J. 1997, ApJ, 482, 355

\bibitem[]{}
Skrutskie, M. F., Cutri, R. M., Stiening, R., \et\ 2006, AJ, 131, 1163

\bibitem[2002]{Sridharan02}
Sridharan, T. K., Beuther, H., Schilke, P., Menten, K. M., \& Wyrowski, F.
2002, ApJ, 566, 931


\bibitem[]{}
Sugitani, K., Fukui, Y., \& Ogura, K. 1991, ApJS, 77, 59


\bibitem[]{}
Schwartz, R. D., Gyulbudaghian, A. L., \& Wilking, B. 1991, ApJ, 370, 263



\bibitem[]{} 
Takeuchi, T., \& Lin, D.\ N.\ C.  2005, ApJ, 627, 286


\bibitem[]{}
Whitney, B. A., Indebetouw, R., Bjorkman, J. E., \& Wood, K. 2004, ApJ, 617, 1177

\bibitem[]{}
Williams, S. J., Fuller, G. A., \& Sridharan, T. K. 2004, A\&A, 417, 115

\bibitem[]{}
Wu, Y., Wei, Y., Zhao, M., \et\ 2004, A\&A, 426, 503

\bibitem[]{}
Wu, Y., Zhang, Q., Chen, H., \et\ 2005, ApJ, 129, 330

\bibitem[]{}
Xu, Y., Shen, Z.-Q., Yang, J., \et\ 2006, AJ, 132, 20

\bibitem[]{} 
Zapata, L. A., Ho, P. T. P., Rodr\'{\i}guez, L. F., O'Dell, C. R., Zhang, Q.,
\et\ 2006, ApJ, 653, 398

\bibitem[]{}
Zhang Q., Hunter, T. R., Brand, J., \et\ 2005, ApJ, 625, 864

\end{thebibliography}
\end{document}